\documentclass[aps,prl,showpacs, twocolumn]{revtex4}
\usepackage{amsmath}
\usepackage{epsfig}
\usepackage{amssymb}
\usepackage{color}
\usepackage{graphicx}

\newcommand{\eqb}{\begin{eqnarray}}
\newcommand{\eqe}{\end{eqnarray}}

\newcommand{\sech}{\mathrm{sech}}

\newcommand{\ZE}{{\textrm{MZE}}}

\def\bes{\begin{subequations}}
\def\ees{\end{subequations}}

\newcommand{\jin}{j_\infty}

\newcommand{\rhin}{\rho_\infty}

\begin{document}

\title{{Macroscopic} Zeno effect and stationary flows in nonlinear waveguides with localized dissipation}

\author{D. A. Zezyulin$^{1}$, V. V. Konotop$^{1,2}$, G. Barontini$^3$, and H. Ott$^3$}
\affiliation{
$^1$Centro de F\'isica Te\'orica e Computacional,   Faculdade de Ci\^encias, Universidade de Lisboa, Avenida Professor Gama Pinto 2, Lisboa 1649-003, Portugal
\\
$^2$ Departamento de F\'isica, Faculdade de Ci\^encias,
Universidade de Lisboa, Campo Grande, Ed. C8, Piso 6, Lisboa
1749-016, Portugal
\\
$^3$ Department of physics and OPTIMAS research center, University of Kaiserslautern, Erwin Schr\"odinger Stra{\ss}e, 67663 Kaiserslautern, Germany
}
\begin{abstract}
We theoretically demonstrate the possibility to observe the
{macroscopic} Zeno effect for nonlinear waveguides with a
localized dissipation. We show the existence of stable stationary
flows, which are balanced by the losses in the dissipative domain.
The {macroscopic} Zeno effect manifests itself in the
non-monotonic dependence of the stationary flow on the strength of
the dissipation. In particular, we highlight the importance of the
parameters of the dissipation to observe the phenomenon. Our
results are applicable to a large variety of systems, including
condensates of atoms or quasi-particles and optical waveguides.
\end{abstract}


\maketitle

{Since the pioneering work of Khalfin concerning the
non-exponential decay of unstable atoms \cite{Khalfin} the
relation between the decay rate and the measurement process was in
the focus of many studies. One of the fundamental results of the
theory, termed after the seminal paper~\cite{Zeno}, as
\emph{quantum Zeno effect}, consists in slowing down the dynamics
of a quantum system subjected to frequent measurements or to a
strong coupling to another quantum system.
 This phenomenon was demonstrated in a rigorous mathematical framework in~\cite{Zeno} and received its further refinements and extension in subsequent studies~\cite{theoretical}.} Experimentally, the quantum Zeno effect has
been confirmed for single ions \cite{Itano}, ultracold atoms in
accelerated optical lattices \cite{Zeno_accel_lattice}, atomic
spin motion controlled by circularly polarized light
\cite{Zeno_spin}, an externally driven mixture of two hyperfine
states of neutral atoms \cite{Zeno_mixture}, photons in a cavity
\cite{Bernu2008} and the production of cold molecular gases in an
optical lattice~\cite{Zeno_molecule}. It has also been predicted \cite{Zoller,KS} that the
tunneling dynamics of particles in a double-well potential can be
slowed down if the particles are removed from one of the wells.
Qualitatively similar results for the suppression of atom losses
in an open Bose-Hubbard chain were reported in~\cite{BoseHubbard}.
In the limit of an infinitely strong measurement of particles in a
given spatial domain it has been shown that the system is projected
onto a unitary dynamics in the loss-free domain~\cite{Facchi2008}.

{The Zeno effect is sometimes also understood in more general
terms as the effects of changing a decay law depending on the
frequency of measurements~\cite{Zeno_Gen}. Applying this
definition to a macroscopic quantum system, like a gas of
condensed bosonic atoms and, taking into account that in the
macroscopic dynamics the frequency of measurement can be
interpreted as the strength of the induced dissipation~\cite{KS},
the effect of the measurement on the decay of the quantum system
can be viewed as the effect of dissipation on the macroscopic
characteristics of the system. Here we assume this interpretation
of the phenomenon and address the questions how the appearance of
localized losses in a waveguide is connected to the appearance of
Zeno-like dynamics. In order to emphasize  the distinction of the
latter statement of the problem with respect to already standard
and widely accepted notion of quantum Zeno effect, below we refer
to {\it macroscopic Zeno effect (MZE)} bearing in mind its
meanfield manifestation.}

Losses are ubiquitous for real quantum systems due to the coupling
to an environment. Very often, the loss processes are also
spatially localized. They can be either externally engineered,
e.g. with the tip of a scanning probe microscope, a local probe in
a quantum gas, an absorbing spatial domain, or they can be intrinsically present in the form of defects
and impurities. One can therefore expect, that the \ZE{}
can manifest itself in a wide class of physical systems, including exciton-polaritons
\cite{excitonpolaritons}, magnon gases \cite{magnons}, surface
plasmons \cite{surfaceplasmons}, and optics of nonlinear Kerr
media~\cite{AKS}.

We study one-dimensional nonlinear waveguides governed by the nonlinear Schr\"odinger equation 
\begin{eqnarray}
\label{GPE} i  \Psi_t=- \Psi_{xx} + g|\Psi|^2\Psi  -
i\gamma(x)\Psi,
\end{eqnarray}
where $g$ the nonlinearity parameter and the local loss processes
are modelled by $i\gamma(x)$ (for a review on the application of
complex potentials see e.g.~\cite{Muga}). {Since the localized
dissipation is applied to a homogeneous condensate, i.e. it breaks
the translational invariance of the system, it can be referred to
as a dissipative defect~\cite{BKPO}}.   We are interested in
\textit{stationary flows}, which correspond to a situation when an
incoming flux of a particles from both ends of the waveguide is
exactly balanced by the losses in the dissipative domain. {Notice
that in such a statement counting of lost particles is replaced by
computing the number of particles that must be loaded into the
system in order to compensate the losses.}

\begin{figure}
\hspace{0.1cm}\includegraphics[width=0.45\columnwidth]{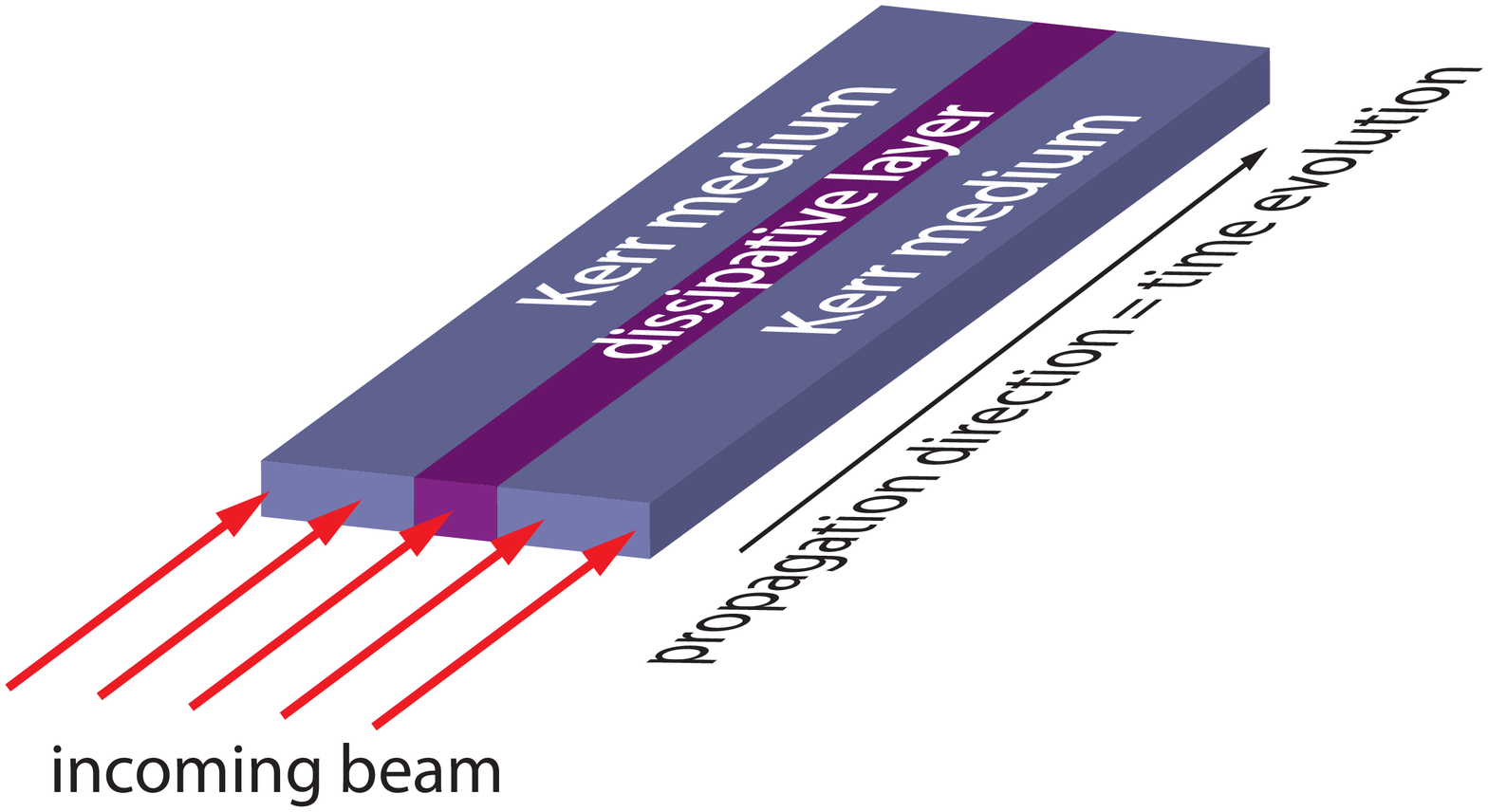}
\hspace{0.5cm}\includegraphics[width=0.45\columnwidth]{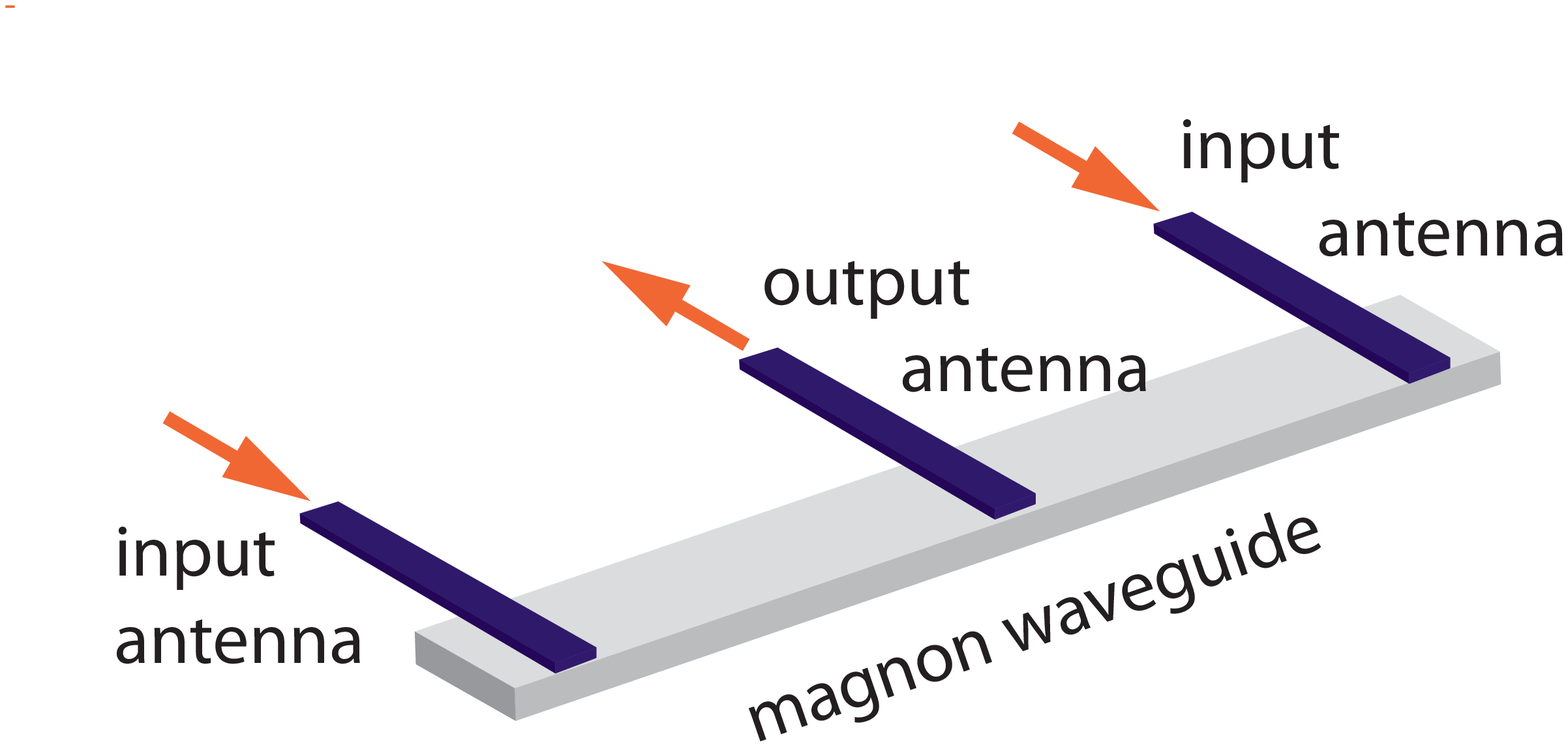}\\
\vspace{1.0cm}
\hspace{0.1cm}\includegraphics[width=0.45\columnwidth]{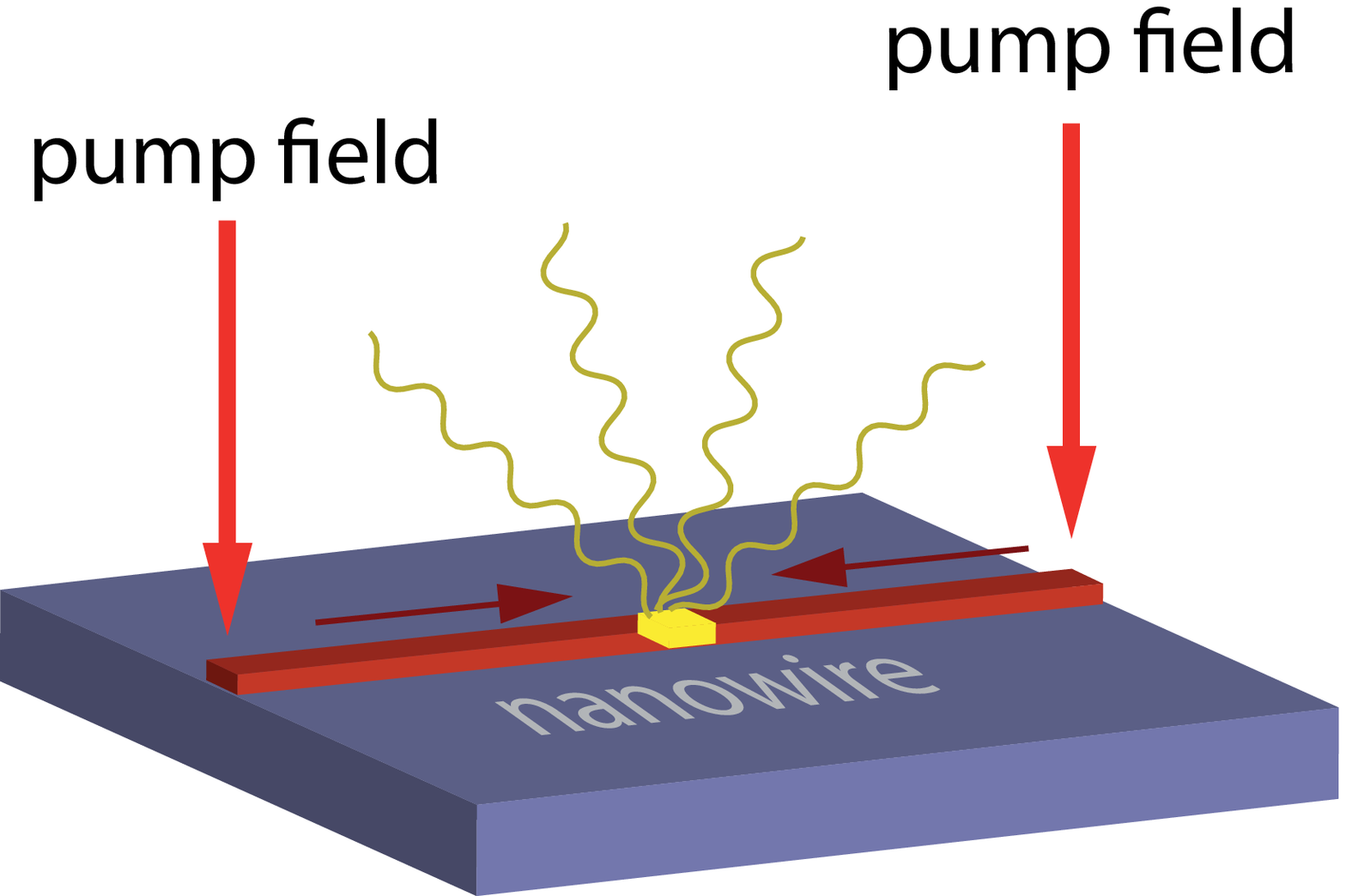}
\hspace{0.5cm}
\includegraphics[width=0.45\columnwidth]{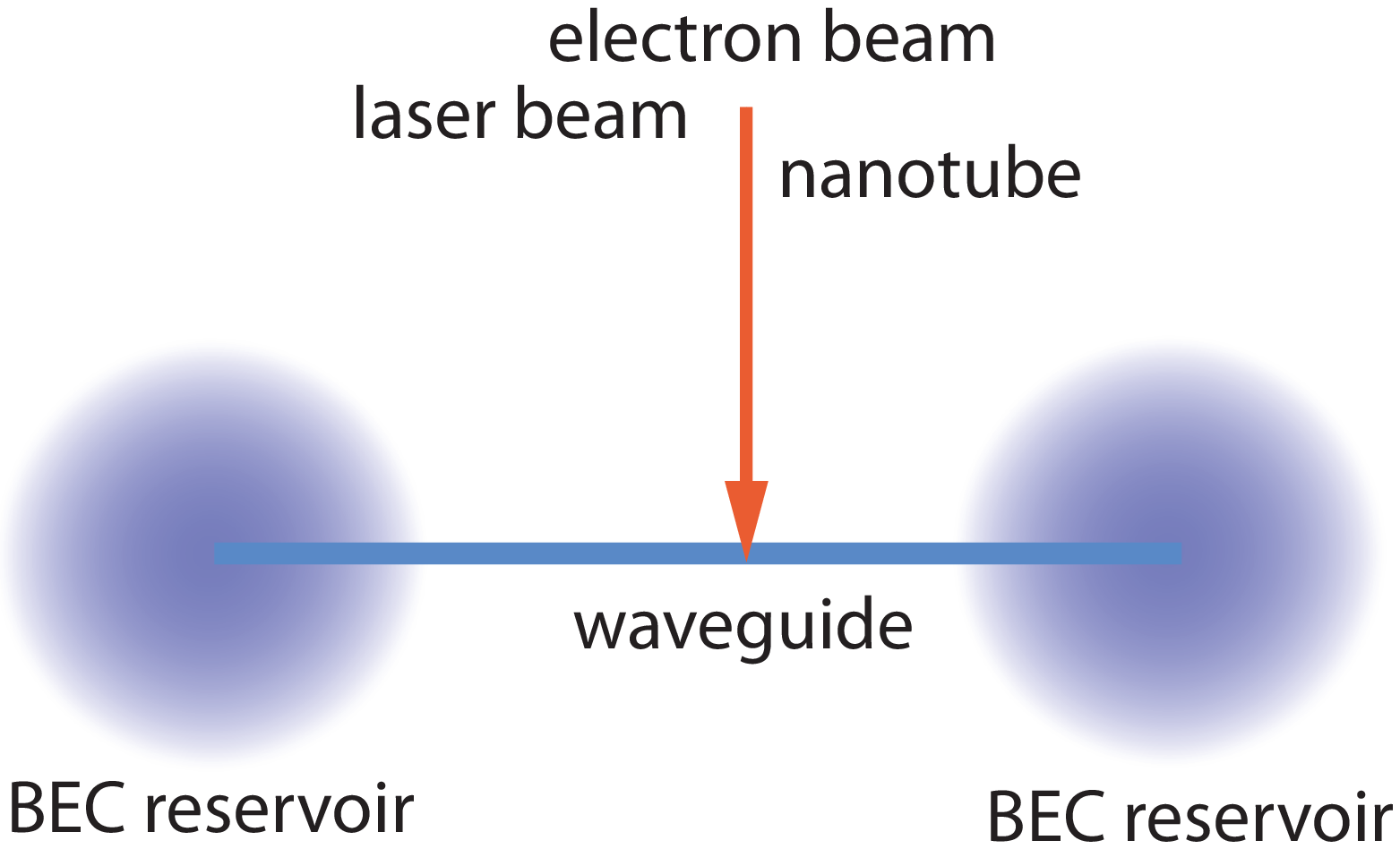}
\caption{Possible experimental scenarios to observe the \ZE:
nonlinear optical waveguide (upper left), a magnon waveguide
(upper right), a plasmonic nanostructure (lower left), an atomic
BEC in a waveguide and two reservoirs (lower right).}
 \label{fig01}
\end{figure}

In Fig.~\ref{fig01} we show several examples of how such a
scenario can be realized experimentally in different physical
systems. While our approach is applicable to a large variety of
physical situations, the systems we have in mind are those
described in \cite{GUHO,Ott,BKPO,Guarrera2011}, i.e. an atomic BEC
subjected to removal of atoms. {In this last case, the time and
the coordinate are respectively measured in units of
$(2\omega_\bot)^{-1}$ and $a_\bot/2$, where $a_\bot$ and
$\omega_\bot$ are the transverse linear oscillator length and the
frequency of the transverse trap, while $g=a_sn_0$, being $a_s$
the scattering length and $n_0$  the unperturbed linear density of
the condensate.} In the following we consider only the case $g>0$
which describes repulsive inter-atomic interactions (or defocusing
Kerr media in optical applications).

The  dissipation is described by the nonnegative  localized
function $\gamma(x)$, which is characterized by two control
parameters: its amplitude $\Gamma_0$ and   characteristic width
$\ell$.  {It is convenient to set $\gamma(x) = \Gamma_0f(x/\ell)$
where $f(x)$ is a known smooth function such that  $\max_x|f(x)| =
f(0)\sim 1$ and   $\max_x|f_x(x)|  \sim 1$.} Then  $\Gamma_0$ and
$\ell$ are proportional  to the intensity of the defect:
$\int_{-\infty}^\infty\gamma(x)dx \propto\Gamma_0\ell$. We will
also assume the most typical experimental situation where
$\gamma(x)$ is an even function: $\gamma(x)=\gamma(-x)$ with only
one maximum at $x=0$. {Having in mind the experiments of
Ref.~\cite{GUHO,Ott,Guarrera2011} one can estimate that
$\Gamma_0\sim I\sigma_{ion}/(e_0\omega_\bot)$ where $I$ is the
current of the electron beam, $e_0$ is the electric charge of the
electron, and $\sigma_{ion}$ is the total ionization cross
section.}

The stationary flows are sought in the form $\Psi_{st}(t,x) = \rho(x)\exp\left[i\int_0^x v(s)ds -i\mu t\right]$, where $v(x)$ is the superfluid velocity, $\mu$ is the chemical potential, and $\rho^2(x)=n(x)$ is the density. Substituting $\Psi_{st}(t,x)$ into Eq.~(\ref{GPE}) we obtain
\begin{eqnarray}
\label{hydro1}
    \rho_{xx} + \mu\rho - g\rho^3 -  j^2  \rho^{-3}=0, \quad
    j_x + \gamma(x)\rho^2 = 0
\end{eqnarray}
with $j(x)= v(x) n(x)$ being the superfluid current. We are interested in solutions of Eqs.~(\ref{hydro1}) with a constant density at infinity:
$\lim_{|x|\to\infty}|\rho(x)|=\rho_\infty$. Then
$\mu = j_\infty^2\rho_\infty^{-4} + g\rho_\infty^2$, where
$j_{\infty} =  \mp \lim_{x\to \pm\infty} j(x)$ is a positive constant. For any stationary flow, the loss of particles in the defect has to be balanced by the incoming current $j_{\infty}$. The main objective of the present study is to show the existence of such stationary flows and to explore the dependence of the  current $j_\infty$ on the parameters of the defect.

First, we consider an example that allows for an exact solution,
extending the result of~\cite{BKPO}. We assume a dissipative
defect of the particular form $\displaystyle{\gamma(x)=
{3\Gamma_0}{\sech^2(x/\ell)}}$. Then it is a straightforward
  to show that $\rho(x)=\tanh(x/\ell)$ and $j(x)= -
\Gamma_0\ell\tanh^3(x/\ell)$ are solutions of the system
(\ref{hydro1}) provided that
$\ell^2(g+\Gamma_0^2\ell^2)=2$.
Thus  the incoming flux  is linearly proportional  to the
intensity of the dissipative defect: $j_{\infty} =
 \Gamma_0\ell$, and in order  to obtain a stationary solution, increase of the strength    of  the dissipation must be compensated with an increase of the incoming flux. In other words, if the incoming flux of particles is increased, the excess particles can only be removed by a stronger defect. While this result is quite
intuitive, we show below that it does not hold in general. In
particular, we will show that for appropriate parameters, an
increasing flux can be compensated by a weaker defect.

We now focus on a dissipation with   finite support:  $\gamma(x) \equiv 0$ if
$|x|>\ell$. This form of the dissipative term models, in
particular, the electronic beam used in~\cite{GUHO,Ott}. In order
to decrease numerical errors we choose $\gamma(x)$ to be smooth at
the edges of the dissipative domain: $\gamma(x) = \Gamma_0 \left(1
- x^2/\ell^2\right)^2$ if $|x|<\ell$.

Since Eq.~(\ref{hydro1}) is not integrable -- unlike its
conservative counterpart where $\gamma(x)\equiv0$ -- it is
convenient to treat $\Gamma_0$ as a parameter that increases
departing from zero. Experimentally, this would correspond to an
adiabatic increase of the defect intensity. For  $\Gamma_0=0$ one
recovers two well-known solutions:   a constant density
$\rho(x)=\rho_\infty$ and a dark soliton {$\rho(x) =
\rho_\infty\tanh(\sqrt{g/2}\,\rho_\infty x)$} [$j(x)\equiv 0$ for
both solutions]. When the defect is adiabatically  switched on,
the constant density and the dark soliton give origin to two
branches of solutions.  The branch bifurcating from the constant
density consists of symmetric flows, for which the relation
$\rho(x)=\rho(-x)$ holds. The flows that  branch off from the dark
soliton are antisymmetric, $\rho(x)=-\rho(-x)$. From the second of
Eqs.~(\ref{hydro1}) it follows that both the symmetric and
antisymmetric flows possess odd currents: $j(x)=-j(-x)$.

Considering behavior of the solutions in the vicinity of $x=0$,
for the symmetric flows  we obtain
$\rho_{xx}(0)=\rho(0)[g(\rho^2(0)-\rhin^2) -
j_{\infty}^2\rhin^{-4}]$. Thus, employing a physically obvious
condition  $\rhin>\rho(0)>0$, we find $\rho_{xx}(0)<0$. We therefore arrive at the counterintuitive conclusion that for the symmetric flows  the atomic density  $n(x)$ has a local maximum in the point of maximal dissipation. 

On the other hand, for $x\to \infty$ both symmetric and
antisymmetric flows behave as
\begin{equation}
\label{eq:asexp}%
\rho_\infty - \rho(x)\propto  e^{-\sqrt{\Lambda}x}, \mbox{ where }
\Lambda = 2g\rho_\infty^2- {4j_\infty^2}{\rho_\infty^{-4}}.
\end{equation}
Thus, for a given density $\rhin$  there exists an upper bound for the maximal current
$j^{max}_\infty=\sqrt{g}\rhin^3/\sqrt{2}$ above which no stationary
flow  can exist.

\begin{figure}
\includegraphics[width=\columnwidth]{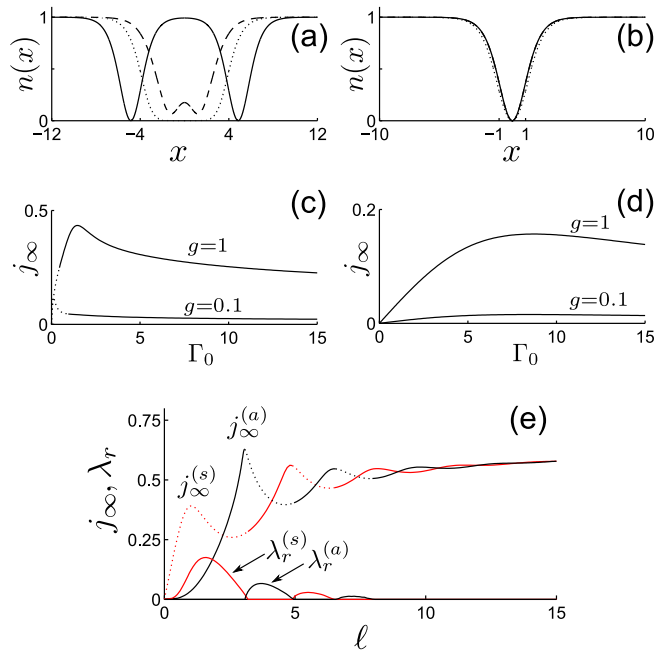}%
  \caption{(a): Density distributions $n(x)$ for  symmetric flows for $g=1$ and $\ell = 4$ and  different values of $\Gamma_0$.  Solid line: $\Gamma_0 = 0.01$,
  dashed line:  $\Gamma_0 = 1$, dotted line:  $\Gamma_0 = 10$.
  (b):   Density distributions $n(x)$ for  antisymmetric flows for $g=1$ and $\ell = 1$.  Solid line: $\Gamma_0 = 0.1$, dashed line:  $\Gamma_0 = 1$, dotted line:  $\Gamma_0 = 10$.
  (c) and (d): Current {\it vs} strength of the dissipation  for symmetric flows (with $\ell=4$)  and for  antisymmetric flows (with $\ell=1$) obtained for   $g= 0.1$   and $g = 1$; stable (unstable) flows correspond to the solid (dotted) fragments of the curves;
  (e):  Currents and instability increments {\it vs} width of the defect for  symmetric [${(s)}$] and antisymmetric  [${(a)}$] flows for  $g=1$ and   $\Gamma_0= 1$. In all panels $\rho_\infty=1$.}%
 \label{fig02}
\end{figure}

In Fig.~\ref{fig02}~(a) [Fig.~\ref{fig02}~(b)] we show the density
profiles $n(x)$  of symmetric [antisymmetric] flows for different values
of the dissipation strength $\Gamma_0$. We observe that  for the symmetric flows, the density  possesses two deep local minima. For a weak dissipation (e.g. $\Gamma_0=0.01$) the  minima are situated outside the dissipative domain, which allows to compute the exact value of the density in these minima: $n_{min} =  {2j_\infty^2}{\rho_\infty^{-4}}/g$. As the strength of the dissipation grows, the  minima  move from $\pm\infty$ towards the center and eventually enter the dissipative domain.
%
%
For antisymmetric flows 
the dependence of the density on $\Gamma_0$  is much weaker pronounced  [the curves for different $\Gamma_0$ are hardly distinguishable on the scale of the Fig.~\ref{fig02}~(b)].

Now our goal is to study the
dependence $\jin$ {\it vs} $\Gamma_0$. The typical results for the symmetric
[antisymmetric] flows are illustrated in Fig.~\ref{fig02}~(c) [Fig.~\ref{fig02}~(d)] . When the nonlinearity is strong enough ($g=1$),  for both types of the flows one can clearly see a global maximum of $\jin(\Gamma_0)$. When $\Gamma_0$ exceeds the value corresponding to this maximum, the required current  $\jin$ decreases. This is the manifestation of the \ZE{}.

In order to observe the \ZE{} in an experiment, it is important that the solutions are stable. To explore the stability of the flows, we substitute $\Psi(t, x) =\Psi_{st}(t, x)+ e^{-i\mu t}\left[a_+(x)e^{\lambda t} + a_-^*(x)e^{\lambda^* t}\right]$ into Eq.~(\ref{GPE}), linearize it
with respect to $a_\pm(x)$, and solve the  obtained linear eigenvalue problem. Instability occurs if there exists an eigenvalue $\lambda$ with   positive real part $\lambda_r$. The results of the analysis are shown in Fig.~\ref{fig02}~(c)--(e). For the symmetric flows presented in Fig.~\ref{fig02}~(c), small values of $\Gamma_0$ do not allow for stable stationary flows (see the dotted fragments of the lines). For instance, the symmetric flow with two local minima situated  outside of the dissipative domain
[shown in Fig.~\ref{fig02}~(a) and corresponding to $\Gamma_0=0.01$]
is unstable. For larger values of $\Gamma_0$ the symmetric flows become stable. We have observed that  the two minima of a stable symmetric flow are  always located in the dissipative domain. All antisymmetric flows shown in Fig.~\ref{fig02}~(d) are stable. Most importantly, the parameter range, in which the \ZE{} is observed, has only stable flows. From experimental point of view, the existence of stable symmetric flows is very appealing, as symmetric flows arise from the overall (symmetric) ground state of the system.

We now elaborate on the role of the size of the defect.
In Fig.~\ref{fig02}~(e), we show the currents {\it vs} $\ell$ for a fixed $\Gamma_0$. As the defect becomes wider, one might
expect a monotonic increase of $j_\infty$. As can be
seen from the graph, this behavior is indeed encountered, however
only in average. Locally it is superimposed by a resonance-like
structure which tends to enhance the current for
certain defect sizes. We also find that  stable solutions only
appear when the $\jin(\ell)$ is a growing function  and that (up
to a certain degree of accuracy) the domains of stability of   the
symmetric flows  coincide with the domains of instability of the
antisymmetric flows  and \textit{vice versa}.

So far, we have encountered two different situations:  in the case
of the $\sech^2$-shaped dissipation, 
the  branch of stationary flows does not show the
\ZE{}. In the case of the dissipation with finite support,
the \ZE{} has various manifestations. The difference can be explained by looking at the asymptotic behavior of the corresponding flows. In  the  case of the $\sech^2$-shaped dissipation,
the asymptotical behavior at $x\to\infty$ of the density is completely
determined by the characteristic width $\ell$ of the defect:
$\rho_\infty -\rho(x) \sim  2e^{-2x/\ell}$, while the
flows supported by  the dissipation with finite support  behave according to Eqs.~(\ref{eq:asexp}). Moreover, Eqs.~(\ref{eq:asexp}) imply that there exists the maximal possible
current $j^{max}_\infty$. It appears that the presence of such a threshold is a signature of the \ZE{}. No such threshold exists for the stationary flows with the $\sech^2$-shaped dissipation: the current $j_\infty$ can be arbitrarily large and the \ZE{} is not found.

However,  the $\sech^2$-shaped dissipation  still does not forbid the
\ZE{} in principle, since flows obeying Eqs.~(\ref{eq:asexp}) can also be found in this case. Let us revisit the dissipation of the form $\gamma(x)=3\Gamma_0\sech^2(x/\ell)$.
Substituting $\rho(x) = \rho_\infty - \rho_1(x)$,  where
$\rho_1(x)=o(1)$ as ${x\to\infty}$, into Eqs.~(\ref{hydro1}) and neglecting the terms of smaller
order, one observes that for $x\gg 1$  the   function  $\rho_1(x)$ is described by the equation $\rho_{1,xx} - \Lambda\rho_1 = 12{j_\infty\Gamma_0\ell}{\rho_\infty^{-1}} e^{-2x/\ell}$.
For $\rho_1(x)$ to obey
the asymptotics (\ref{eq:asexp}) the two conditions must be fulfilled:
(i) $\Lambda>0$ -- gives the
\textit{maximal} current: $j_\infty < j^{max}_\infty$; (ii) $\sqrt{\Lambda} < 2/\ell$
-- yields the \textit{minimal} possible
current: $j_\infty>j^{min}_\infty= \rho_\infty \sqrt{
g\rho_\infty^4/2-\ell^{-2}}$ (if the expression under the radical is negative, then $j^{min}_\infty=0$). As $\ell$ grows, $j^{min}_\infty$  approaches $j^{max}_\infty$. Hence, the  range of currents allowing for  the solutions  which obey Eqs.~(\ref{eq:asexp}) decreases.  This leads us to the conjecture that
\textit{rapidly decaying dissipation is  favorable for the observation of the macroscopic Zeno effect}. In particular,  the defects decaying faster than exponentially are more likely to display the \ZE{} than  ones obeying exponential decaying.

We close this Letter with a discussion of   possible experimental observation of the found \ZE{}. The incoming flux of particles has to be generated at both ends of the waveguide. This can be achieved by controlled pumping terms in the case of quasiparticles or by reservoirs in the case of real particles. For light propagating in a non-linear waveguide, such boundary conditions appear rather naturally. But even for a finite system with no reservoir, one can speculate that a quasi stationary state is established on intermediate time scales in a transient regime: if the defect is switched on in a finite system that is initially in its ground state, the stationary flow will develop out of symmetric initial conditions and will retain its symmetry with increasing dissipation. With time, a flow of particles towards the defect is created which can mimic the boundary conditions, applied in the preceding discussion. The condition of having a defect that drops faster than exponentially can be easily realized in most of the experimental implementations suggested above.

\begin{figure}
\includegraphics[width=\columnwidth]{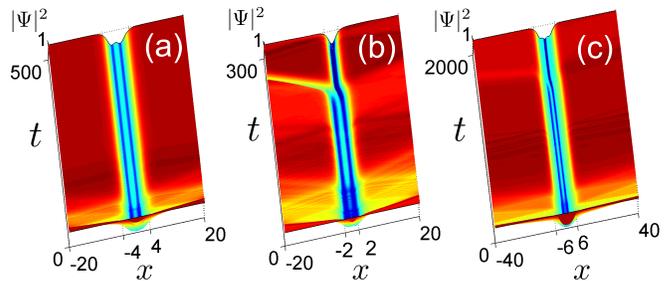}%
  \caption{Evolution of  the density $|\Psi(t,x)|^2$  starting from the  initial data $\Psi(0, x) = 1$.
   For all the shown panels $\rho_\infty = g = \Gamma_0= 1$.
   (a): The width of the defect is $\ell=4$. The generation of the symmetric stationary flow occurs.
   (b) and (c): The width of the defect is  $\ell=2$ and $\ell=6$ respectively.  The symmetric flows are unstable, and therefore no stationary flow is established.}%
 \label{fig03}
\end{figure}

We now support this reasoning by illustrating the generation of stationary flows
through direct integration of Eq.~(\ref{GPE}) on a finite domain subject to the boundary conditions
$\Psi(t, \pm L) = \rho_\infty e^{-i\mu t}$  (here $L$ is the  half-width of the computational domain).
In Fig.~\ref{fig03} we show the temporal evolution of the atomic density for three different widths of the defect. For all the shown evolutions, the initial density is taken to be constant and the chosen boundary conditions fix the density and the chemical potential at the edges of the computational domain.  
Figure~\ref{fig03}~(a) shows the  evolution for a set of
parameters, where a stable symmetric stationary flow exists. After
initial decrease of the density at the location of the defect, the
system achieves the stationary flow. In the vicinity of the origin
one clearly observes the two local minima residing inside the
defect. Indeed, the density approaches eventually the stationary
flow profile not only in the vicinity of the origin but in the
entire computational domain. However, if the symmetric flow is not
expected to be stable [Fig.~\ref{fig03}~(b) and
Fig.~\ref{fig03}~(c)], then the density profile  shows ongoing
distortions in the vicinity  of the origin and eventually loses
the symmetry. For the cases shown in panels (b) and (c),  the
density tends to approach a profile corresponding to a stable
antisymmetric flow which exists for the  chosen values of  $\ell$
and $\Gamma_0$ [recall that the domains of the stability of the
symmetric and antisymmetric flows alternate as shown in
Fig.~\ref{fig02}~(e)]. However, in the cases  (b) and (c) truly
stationary antisymmetric flows are not established because the
chosen boundary conditions can support only symmetric flows, which
are unstable for the chosen parameters.

In summary, we have analyzed a non-linear waveguide with a
localized dissipative defect. We found evidence for the appearance
of the  {macroscopic} Zeno effect for specific boundary conditions
and have analyzed to role of the interaction and the defect size.
The observed {macroscopic} Zeno effect is intrinsically related to
the existence of the stable stationary flows and is expressed
through nonmonotonic dependence of the intensity of the
dissipation on the density of incoming currents. The proofed
existence of stable solutions for symmetric flows is very
important, as these solutions correspond to many natural
experimental situations.

\begin{acknowledgements}
The collaborative research was supported by {the
bilateral program between} DAAD (Germany) and FCT (Portugal). We
acknowledge financial support by the DFG within the SFB/TRR 49,
and FCT through the grant PEst-OE/FIS/UI0618/2011. DAZ is
supported by FCT under the grant  No. SFRH/BPD/64835/2009.
GB is supported by a Marie Curie Intra-European Fellowship.
\end{acknowledgements}

\end{document}